\documentclass{gENO2e}

\usepackage{graphicx}
\usepackage{natbib}
\usepackage[T1]{fontenc}
\renewcommand{\baselinestretch}{1} 
\begin{document}

\title{Simulated Annealing for Weighted Polygon Packing}     
  
\author{Yi-Chun Xu$^{a}$, Ren-Bin Xiao$^{b}$ and Martyn Amos$^{c,\dagger}$\thanks{$\dagger$ Email: M.Amos@mmu.ac.uk} \\ \vspace{6pt} $^{a}$ {\it Institute of Intelligent Vision \& Image Information, China Three Gorges University, China} \\ $^{b}$ {\it Institute of System Engineering, Huazhong University of Science \& Technology, China} \\ $^{c}$ {\it Department of Computing \& Mathematics, Manchester Metropolitan University, UK}}


\maketitle

\begin{abstract}  
In this paper we present a new algorithm for a layout optimization problem: this concerns the placement of
weighted polygons inside a circular container, the two objectives
being to minimize imbalance of mass and to minimize the radius of
the container. This problem carries real practical significance in
industrial applications (such as the design of satellites), as well
as being of significant theoretical interest. Previous work has dealt with circular or rectangular objects, but here we deal with the more realistic case where objects may be represented as polygons and the polygons are allowed to rotate. We present a solution based on simulated annealing and first test it on instances with known optima. Our results show that the algorithm obtains container radii that are close to optimal. We also compare our method  with existing algorithms for the (special) rectangular case.
Experimental results show that our approach out-performs these methods in terms of solution quality.
\end{abstract}
%
\section{Introduction}

The {\it Layout Optimization Problem} (LOP) concerns the physical
placement of instruments or pieces of equipment in a spacecraft or
satellite. Because these objects have mass, the system is subject to
additional constraints (beyond simple Cartesian packing) that affect
our solution. The two main constraints that we handle
in this paper are (1) the space occupied by a given collection of
objects (envelopment), and (2) the non-equilibrium (i.e. imbalance) of the system.
The rest of the paper is organized as follows: 
In Section 2 we first present a
detailed description of the problem, and describe previous related
research. In Section 3 
we describe our algorithm, and in Section 4 we give the results of numerical experiments.

\section{The Layout Optimization Problem in Satellites}

The Layout Optimization Problem (LOP) was proposed by Feng {\it et al.} \cite{layout:thf} in 1999, and has significant implications for the cost and performance of devices such as satellites and spacecraft. It concerns the {\it two dimensional} physical placement of a collection of {\it objects} (instruments or other pieces of equipment) within a spacecraft/satellite ``cabinet", or {\it container}. The LOP is demonstrably NP-hard \cite{layout:NPhard}. Early work on this problem \cite{layout:lining,layout:tangfei,layout:yuyang,layout:zhouchi} almost always modeled objects as circles in order to simplify the packing process. However, in real-world applications, objects are generally rectangular or polygonal in shape, and modeling them as circles leads to expensive wastage of space. We have recently reported work on solving the rectangular case \cite{icnc07}, and here we report a new algorithm (based on a different approach) to solve the polygonal case.

We now briefly introduce related work on the packing of irregular items. Dowsland {\it et al.} use a so-called "Bottom-Left" strategy to place polygonal items in a bin \cite{Dowsland02}, with items having fixed orientations. Poshyanonda {\it et al.} \cite {Poshyanonda02} combine genetic algorithms with artificial
neural networks to obtain reasonable packing densities. In other related work, Bergin {\it et al.} study the packing of identical rectangles in an irregular container \cite{BirginCOR06,BirginORS06}. Burke and Kendall have applied simulated annealing to translational polygon packing (i.e., without rotations) \cite{Burke97}. Other authors have applied simulated annealing to solve the problem of rotational polygon packing on a continuous space \cite{heckmann1995saa,Jain92}.

Given the additional constraint that imbalance of mass must be minimised, it is difficult to see how these existing methods may be directly applied to the current problem. In what follows we describe a new nonlinear optimization model for the LOP, and then show how it may be solved using simulated annealing.

\subsection{Notation and Definitions}

Here we describe the formal optimization model, by first explaining our notation for the representation of polygons. We then show how to quantify relations between polygons (such as distance and degree of overlap), which are central to the problem of assessing the overall quality of a layout.

\subsubsection{Structure of a polygon}

\begin{figure} 
\begin{centering}
\includegraphics[scale=0.8]{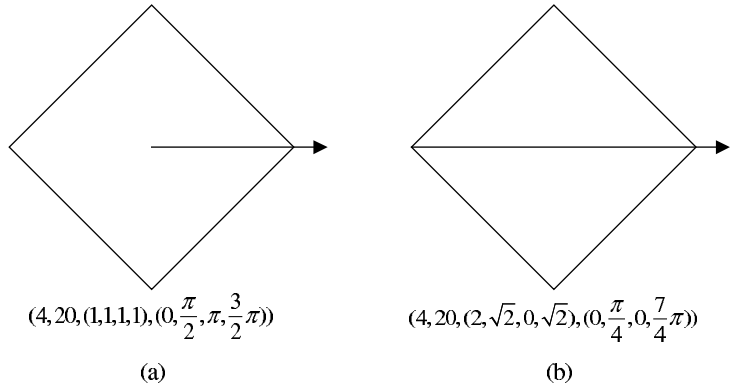}
\caption{Illustration of the structure definition of polygons}
\label{fig_structures}
\end{centering}
\end{figure}

Suppose there are $k$ polygons $(1, 2, \dots, k)$  to be packed. The {\it structure} of a polygon includes both its {\it shape} and its {\it mass}. We use $str(i)$ to denote the initial structure of a polygon, $i$:

\begin{equation}
str(i) =(n_i,m_i,(r_1,r_2,..., r_{n_i}),(\theta _1, \theta _2, ..., \theta _{n_i}) )
\end{equation}

where $m_i$ is the mass of polygon $i$, and $n_i$ is the number of vertices in the graph representation of polygon $i$. The positions of the $n_i$ vertices are defined by two lists of {\it polar} coordinates. List $(r_1,r_2,..., r_{n_i})$ defines the {\it Euclidean distance} from each of the $n_i$ vertices to the polygon's centre of mass, and list $(\theta _1, \theta _2, ..., \theta _{n_i})$ defines the {\it orientation} of each of the $n_i$ vertices relative to the centre of mass. Figure ~\ref{fig_structures} shows how to define the initial structure of a square with edge length 1; in Figure ~\ref{fig_structures} (a), the shape's centre of mass is located at the shape centre, whereas in  Figure ~\ref{fig_structures}(b), the centre of mass is located at one vertex. We define the point of reference of each polygon as its center of mass, in order to simplify the notation.

\subsubsection{Radius of a polygon}

The radius of polygon $i$ is defined as the maximum of $(r_1,r_2,..., r_{n_i})$:

\begin{equation}
r(i)=\max \{r_1, r_2, ..., r_{n_i}\}
\end{equation}

With the polygon's centre of mass at its own centre, the circle with radius $r(i)$ defines the minimum-sized circle that can completely cover the polygon.
   
\subsubsection{State of a polygon}

\begin{figure} 
\begin{centering}
\includegraphics[scale=0.8]{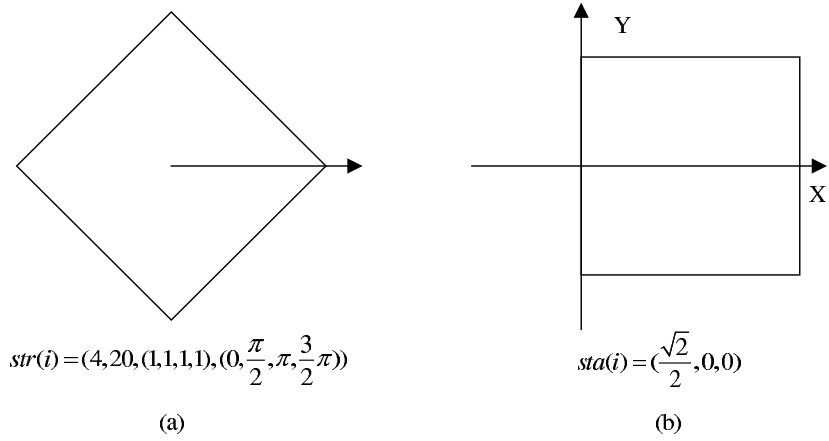}
\caption{Illustration of the state of a polygon}
\label{fig_states}
\end{centering}
\end{figure}

We use Cartesian coordinates to record the positions of the polygons, and set the center of the container (that is, the circle) as the original point. We use $sta(i)$ to denote the state of a polygon $i$:

\begin{equation}
sta(i) =(x_i,y_i,\alpha _i )
\end{equation}

where $x_i, y_i$ is the position of the centre of mass, and $\alpha$ defines a rotation angle. Then with $str(i)$ and $sta(i)$, we can draw a polygon, $i$, as in Figure ~\ref{fig_states}.

\subsubsection{Distance between two polygons}

The distance between two polygons is defined as the Euclidean distance between their centres of mass:

\begin{equation}
dis(i,j)=\sqrt{(x_i-x_j)^2+(y_i-y_j)^2)}
\end{equation}

\subsubsection {Overlap between two polygons}

If two polygons do not overlap, this measure is zero. If two polygons $i$ and $j$ overlap at all, we measure this as:

\begin{equation}
\label{equ:overlap}
 ove(i,j)= \max \{0, r(i)+r(j)-dis(i,j) \}
\end{equation}

\begin{figure} 
\begin{centering}
\includegraphics[scale=1]{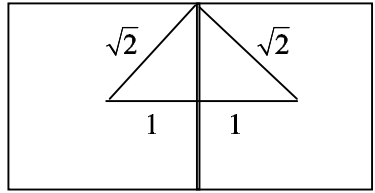}
\caption{The overlap function is not continuous}
\label{fig_overlaps}
\end{centering}
\end{figure}

\noindent
This measurement of overlap has certain characteristics:

\begin{itemize}
\item In Equation (\ref{equ:overlap}), $r(i)$ and $r(j)$ are two constants, therefore $dis(i,j)$ can be directly obtained by the positions of the two polygons.
\item It is clear that $ove(i,j) \geq 0$. To satisfy the {\it non-overlapping} constraint, we should minimize $ove(i,j)$ to zero.
\item $ove(i,j)$ is not a continuous function of the positions. As shown in Figure  ~\ref{fig_overlaps}, when two squares with edge length 2 are adjacent on one side, their overlap is zero, but when the left square is moved a little to the right, the overlap ``jumps" to $2\sqrt2-2$.  
\end{itemize}

Ascertaining overlap between two polygons is not a difficult problem in computational geometry or computer graphics. In this paper, we look at each edge of one polygon in turn; if it is intersected by any edge of another polygon, then an overlap exists; otherwise, if one polygon is contained {\it within} another, then clearly an overlap exists. So the ascertaining of overlap has complexity $O(mn)$, where $m,n$ are the number of edges of the two polygons. 

\subsubsection{State of a layout}

A layout $X$ is defined as the combination of the states of $k$ polygons:

\begin{equation}
X=(x_1, y_1, \alpha _1, x_2, y_2, \alpha _2, ..., x_k, y_k, \alpha _k)
\end{equation}

\subsubsection{Radius of a layout}

If $(X_x, X_y)$ denotes the position of the centre of mass of a layout $X$, then

\begin{equation}
X_x= \frac{\sum_{i=1}^{k}m_ix_i}{\sum_{i=1}^{k}m_i},
X_y= \frac{\sum_{i=1}^{k}m_iy_i}{\sum_{i=1}^{k}m_i}.
\end{equation}

We define the radius $r(X)$ of a layout $X$ as the longest Euclidean distance from its centre of mass to any of the vertices of the polygons.
Because of the {\it imbalance} constraint, we place the centre of mass of the layout at the container center, so $r(X)$ defines the minimum-sized container.

\subsubsection{Overlap of a layout } 

The overlap of a layout is the sum of all overlaps between its polygons:

\begin{equation}
ove(X)=\sum\limits_{i=1}^{k}\sum\limits_{j=1,j\not =i}^{k} ove(i,j)
\end{equation}

\subsubsection{Problem definition}

From the definitions above, we obtain an unconstrained optimization problem:

 \begin{equation}
 \mbox{minimize } f(X)=\lambda _1 ove(X)+\lambda _2r(X)
 \label{eq_energy}
\end{equation}

where $\lambda_ 1,\lambda_ 2$ are two constants. Because the overlap function $ove(X)$ is not continuous, $f(X)$ is not continuous. In general, the overlaps are extremely deleterious, so $\lambda _1$ should be set large enough to prevent their introduction. However, we note that the computation does not introduce overlaps when attempting to decrease the radius of the layout at the final stage of optimization, because of the discontinuous of the $ove$ function. 

\section{Simulated annealing algorithm}

{\it Simulated annealing} (SA) is a probabilistic meta-heuristic that is well-suited to global optimization problems \cite{Kirkpatrick83}. {\it Annealing} refers to the process of heating then slowly cooling a material until it reaches a stable state. The heating enables the material to achieve {\it higher} internal energy states, while the slow cooling allows the material more opportunity to find an internal energy state {\it lower} than the initial state.  SA models this process for the purposes of optimization. A point in the search space is regarded as a system state, and the objective function is regarded as the internal energy. Starting from an initial state, the system is perturbed at random, moving to a new state in the neighbourhood, and a change of energy $\Delta E$ takes place. If $\Delta E<$0, the new state is accepted (a {\it downhill} move), otherwise the new state is accepted with a probability $\exp (\frac{-\Delta E}{K_b T})$ (an {\it uphill} move), where $T$ is the temperature at that time and $K_{b }$ is Boltzmann's constant. When the system reaches equilibrium, $T$ is decreased. When the temperature approaches zero, the probability of an ``uphill" move  becomes very small, and SA terminates.

Let $t_0$ denote the initial temperature, $imax$ denote the maximum number of iterations, $E(x)$ denote the energy function, and $emax$ denote the "stopping" energy.The general pseudo-code for simulated annealing may be written as follows: 

\begin{tabbing}
=====\===\===\===\===\===\ \kill
\> {\bf Algorithm 1}: Standard SA\\
\\
\> set the initial state $x$ and initial temperature $t=t_0$, let $i=0$\\
\> {\bf while} $i<imax$ {\bf and} $E(x)> emax$  \\
\> \>  \>  perturb $x$ in its neighbourhood and get $x'$ \\
\> \>  \>  {\bf if} $E(x')<E(x)$ {\bf then} $x=x'$ \\
\> \>  \>  {\bf else} $x=x'$ with probability $\exp (\frac{E(x)-E({x}')}{t})$ \\
\> \>  \>  decrease the temperature \\
\> \>  \>  $i=i+1$\\
\> \bf{return} $x$\\
\end{tabbing}

The performance of SA may be affected by several parameters: the initial temperature $t_0$, the maximum number of iterations, $imax$, the ``stopping" energy, $emax$, the structure of the neighbourhood, and the schedule of cooling. For a given problem, the values of these parameters should be carefully selected.

\subsection{SA for polygon packing}

\subsubsection{Neighbourhood structure}

In each iteration of SA, we perturb one polygon, thus obtaining a new layout, and then decide whether to accept or reject the new layout by means of an evaluation. In 
the $i$th iteration, the $(i~mod~ k)$th polygon will be perturbed.  Given an initial radius $R_0$, which is large enough to contain the polygons, the neighbourhood for polygon $j$ is defined as:

 \begin{equation}
x_j,y_j \in (\frac{imax}{i-2\times imax}+1.05)\times R_0\times random(-1,1)
\label{eq_xyneighbor}
\end{equation}
\begin{equation}
\alpha _j \in(\frac{imax}{i-2\times imax}+1.05) \times \pi \times random(-1,1) 
\label{eq_aneighbor} 
\end{equation}

Equations (\ref{eq_xyneighbor}) and (\ref{eq_aneighbor}) show that, at the beginning of the algorithm's execution, the position of a polygon may vary by ($-0.55 R_0, 0.55 R_0)$, and its orientation may be perturbed by ($-0.55 \pi, 0.55 \pi$). This neighbourhood is large, and and the polygon can thus ``explore" more space. As the algorithm proceeds, the neighbourhood becomes increasingly smaller. At the end of the algorithm's execution, the neighbourhood shrinks to 0.05 times its original size, then SA chooses the best solution in the neighbourhood.

\subsubsection{Temperature decreasing}

We use a simple rule to decrease the temperature: every $cmax$ iterations, we let $t=d \times t$, where $d < 1$. 

\subsubsection{Description of the algorithm}

The detailed SA algorithm for polygon packing is therefore described as follows:

\begin{tabbing}
=====\===\===\===\===\===\ \kill
\>  {\bf Algorithm 2}: SA for the packing problem\\

\\
\> randomly generate an initial layout $X$. Let $t=t_0$, $i=0$\\
\> {\bf while} $i<imax$ {\bf and} $f(X)> emax$  \\
\> \>  \>  let $j = (i~mod~k)$\\
\> \>  \>  randomly select $x_j, y_j, \alpha _j$ by (\ref{eq_xyneighbor}) and (\ref{eq_aneighbor}) and get new $X'$ \\
\> \>  \>  {\bf if} $f(X')<f(X)$ {\bf then} $X=X'$ \\
\> \>  \>  {\bf else} $X=X'$ with probability $\exp (\frac{f(X)-f({X}')}{t})$ \\
\> \>  \>  {\bf if} $i~mod~cmax = cmax-1$ {\bf then}~$t=d \times t$  \\
\> \>  \>  $i=i+1$\\
\> \bf{return} $x$\\
\end{tabbing}

\section{Numerical Results}

We are not aware of any standard library of benchmark instances for this {\it particular} problem, although such libraries do exist for other related problems \cite{fekete2007pil}. We therefore take a two-stage approach to testing our algorithm; we first design six instances with {\it known} optima, against which we may initially validate our method. We note that these instances include both convex and nonconvex polygons. After establishing the effectiveness of our SA algorithm, we then test our algorithm against other recently-described methods for {\it rectangle} packing (rectangles, of course, being members of the polygon class), using both existing instances from the literature and new, larger instances.

\subsection{Known Optima}

\begin{figure}[h]
\begin{centering}
\includegraphics[scale=0.80]{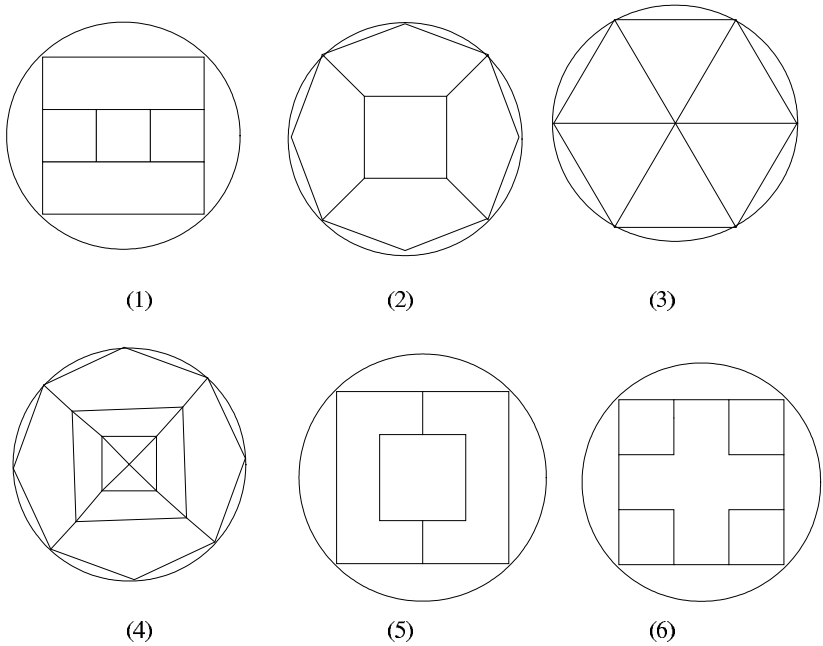}
\caption{Instances with known optima.}
\label{fig_optimal_instances}
\end{centering}
\end{figure}

The instances with known optima are described in Table \ref {table_optimal_instances}, with graphical representations given in Figure \ref{fig_optimal_instances}.

\begin{table}
\begin{centering}
\caption{Four instances with known optima}
\begin{tabular}{@{}c c c l@{}}
\hline
Instance & $k$ & $R_0$ & Structure \\
\hline
 1 & 5 & 2.3 & $str(i)=(4, 30, (\frac{\sqrt{10}}{2}, \frac{\sqrt{10}}{2}, \frac{\sqrt{10}}{2}, \frac{\sqrt{10}}{2})$, \\
&&& $(atan(\frac{1}{3}), \pi-atan(\frac{1}{3}),  \pi+atan(\frac{1}{3}), 2\pi-atan(\frac{1}{3})), i=1 ,2$\\
           &&&         $str(i)=(4, 10, (\frac{\sqrt2}{2}, \frac{\sqrt2}{2}, \frac{\sqrt2}{2},\frac{\sqrt2}{2}), (\frac{\pi}{4}, \frac{3}{4}\pi,  \frac{5}{4}\pi, \frac{7}{4}\pi), i=3,4,5$\\
\hline
 2 & 5 & 2.8 & $str(i)=(5, 100, (2, 2\sqrt2-2, 2, \sqrt2, \sqrt2),$ \\
&&& $(0, \frac{1}{4}\pi,     \frac{1}{2}\pi, \frac{5}{4}\pi, \frac{7}{4}\pi)), i=1,2,3,4$, \\
 &&&                    $str(i)=(4,100,(\sqrt2, \sqrt2, \sqrt2, \sqrt2),(0, \frac{1}{4}\pi, \frac{1}{2}\pi, \frac{3}{2}\pi)), i=5$ \\ 
\hline
 3 & 6 & 3.4 & $str(i)=(3, 100, (2, 2, 2), (0, \frac{2}{3}\pi,  \frac{4}{3}\pi ), i=1,2,3,4,5,6$\\

\hline
 4 & 12 &5.0 & $str(i)=(3,10,(1,1,1),(0, \pi,  \frac{3}{2}\pi )),i=1,2,3,4$,\\
     &&&                  $str(i)=(4,20,(2, 2, \sqrt2, \sqrt2),(0, \pi, \frac{5}{4}\pi, \frac{7}{4}\pi)),i=5,6,7,8$,\\
     &&&                  $str(i)=(4,20,(2, 2, \sqrt2, \sqrt2),(0, \pi, \frac{5}{4}\pi, \frac{7}{4}\pi),i=9,10,11,12$ \\

\hline
 5 & 3 & 8.0 & $str(i)=(4,40,(2\sqrt2, ,2\sqrt2, \sqrt2,2\sqrt2 ),(\frac{1}{4}\pi, \frac{3}{4}\pi ,\frac{5}{4}\pi ,\frac{7}{4}\pi)), i=1$,\\
     &&&      $str(i)=(8, 60, (2\sqrt2, 2\sqrt5,2\sqrt5,2\sqrt5,2\sqrt5,2\sqrt2,2,2), (\frac{1}{4}\pi, atan(2),$ \\
  &&&  $\pi-atan(2), \pi+atan(2), -atan(2), -\frac{1}{4}\pi, -\frac{1}{2}\pi,\frac{1}{2}\pi)),i=2,3$\\
\hline

6 & 5 & 5.0 & $str(i)=(4,60,(\sqrt2,\sqrt2,\sqrt2,\sqrt2),(\frac{1}{4}\pi,\frac{3}{4}\pi,\frac{5}{4}\pi,\frac{7}{4}\pi)),i=1,2,3,4$\\
&&& $str(i)=(12,500,(\sqrt{10},\sqrt{10},\sqrt2,\sqrt{10},\sqrt{10},\sqrt2,$\\
&&& $\sqrt{10},\sqrt{10},\sqrt2,\sqrt{10},\sqrt{10},\sqrt2),(-atan(\frac{1}{3}),atan(\frac{1}{3}),$\\
&&&$\frac{1}{4}\pi, \frac{1}{2}\pi-atan(\frac{1}{3}), \frac{1}{2}\pi+atan(\frac{1}{3}),  \frac{3}{4}\pi,\pi-atan(\frac{1}{3}),\pi+atan(\frac{1}{3}),$\\
&&& $\frac{5}{4}\pi,\frac{3}{2}\pi-atan(\frac{1}{3}),\frac{3}{2}\pi+atan(\frac{1}{3}),\frac{7}{4}\pi)),i=5$\\
\hline
\end{tabular}
\label{table_optimal_instances}
\end{centering}
\end{table}

For each instance, we use SA to try to find the optimal layout. The value of $imax$ is set to $20000\times k$, the value of $cmax$ is set to $100 \times k$, and the initial temperature is set to 100. The constants $\lambda _1$ and $\lambda _2$ are set to 100 (to induce a large $f(x)$ and adjust the probability of uphill movement). In each case, the algorithm is executed 40 times.

\begin{figure}[h]
\begin{centering}
\includegraphics[scale=0.8]{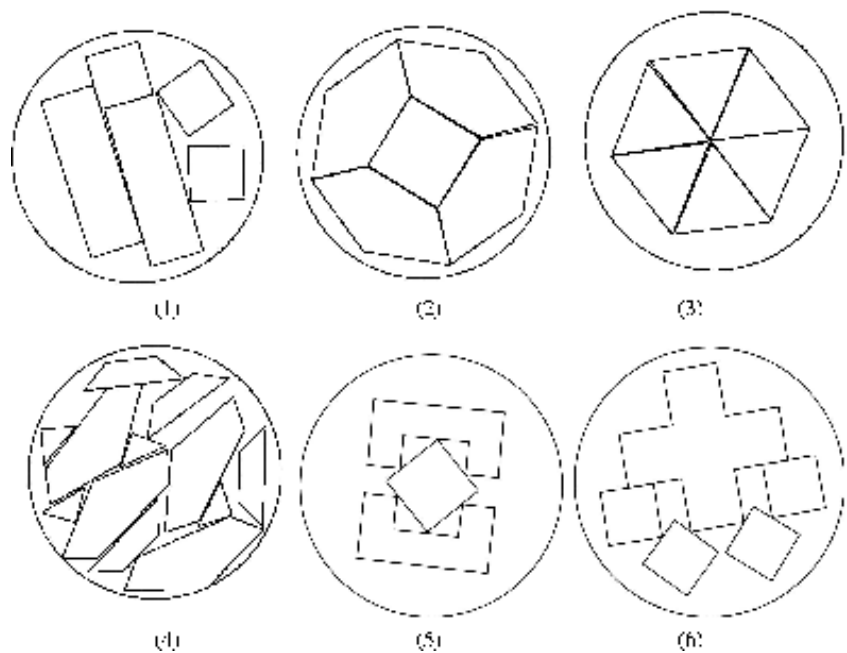}
\caption{Best results obtained for instances with known optima}
\label{fig_layouts}
\end{centering}
\end{figure}

Values for the best radius found, $r_{best}$, mean radius, $\bar r$, and variance $v$ are presented in Table \ref{table_results}. Representations of the best results obtained are given in Figure \ref{fig_layouts}. From Table \ref{table_results} and Figure \ref{fig_layouts}, we observe that the SA algorithm can find layouts that are very close to the the optimal configuration for instances 1, 2, and 3, where the number of polygons is relatively small and the overall structures are simple. The optimal radius for instance 1 was originally calculated as $\frac{3}{2} \sqrt{2}=2.121$, but the our results yielded a smaller value. This prompted a re-estimation, giving a new optimum of $\sqrt{4\frac{9}{64}}$). In the first three instances, the errors to the optimum are about  $\frac {r_{best}-r_{optimum}}{r_{optimum}}=2\%$. The algorithm performs less well on instance 4, with 12 polygons in a relatively complex configuration. In this case, our algorithm cannot find the best configuration, and the error to the optimum is 15\%. Instances 5 and 6 feature noncovex polygons. Because of the complexity of the shapes, the algorithm is unable to solve these instances to optimality. The errors to the optimal radius are 11\% and 4\% for instances 5 and 6 respectively.

\begin{table}[h]
\begin{centering}
\caption{Numerical results for SA on four instances with known optima}
\begin{tabular}{c c c c c}
\hline
      Instance  & Est. $r_{optimum}$ & $r_{best}$  & $\overline{r}$ & $v$  \\ \hline
        1       & $\sqrt{4\frac{9}{64}}=2.034$  & 2.080       & 2.167    & 0.027\\
        2       & $2\sqrt2=2.828$           & 2.861       & 3.209    & 0.010\\
        3       & $2\sqrt3=3.464$           & 3.522       & 4.065    & 0.157\\
        4       & $3\sqrt2=4.242$           & 4.887       & 5.149    & 0.024\\
        5       & $4\sqrt2=5.656$           & 6.295       & 9.266    & 59.25\\
        6       & $3\sqrt2=4.242$           & 4.423       & 7.034    & 119.44\\
\hline    
\end{tabular}
\label{table_results}
\end{centering}
\end{table}

\subsection{Rectangular Instances}

\begin{figure}[h]
\begin{centering}
\includegraphics[scale=0.9]{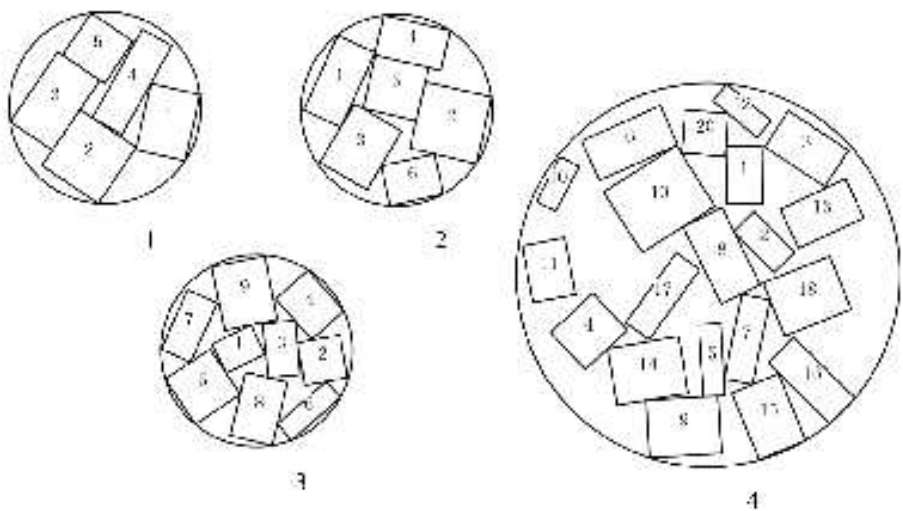}
\caption{Four instances of the LOP using rectangles}
\label{fig_instances_literature}
\end{centering}
\end{figure}

We now test our algorithm on four instances of the LOP containing only rectangular shapes. The first three instances were first described in \cite{zhai1999}, and the fourth in \cite{icnc07}, where all four instances were used to benchmark three different approaches: a genetic algorithm (GA), 
particle swarm optimisation (PSO) and a hybrid compaction algorithm followed by particle swarm local search (CA-PSLS). Depictions of each instance are depicted in Figure ~\ref{fig_instances_literature}, and full descriptions are given in Table ~\ref{table_literature_instances}. Since both particle-based algorithms out-performed the GA, we do not consider this last method here.

\begin{table} 
\begin{centering}
\caption{Four instances from the literature}
{\tiny 
\begin{tabular}{@{}c c c l@{}}
\hline
Instance & $k$ & $R_0$ & Structure \\
\hline
1& 5 & 20& $str(1)=(4,12,(\sqrt{25},\sqrt{25},\sqrt{25},\sqrt{25}),(atan(\frac{3}{4})),\pi-atan(\frac{3}{4}),\pi+atan(\frac{3}{4}), 2\pi-atan(\frac{3}{4})))$\\
&&& $str(2)=(4,16,(\sqrt{32},\sqrt{32},\sqrt{32},\sqrt{32}),(atan(\frac{4}{4})),\pi-atan(\frac{4}{4}),\pi+atan(\frac{4}{4}), 2\pi-atan(\frac{4}{4})))$\\
&&& $str(3)=(4,15,(\sqrt{34},\sqrt{34},\sqrt{34},\sqrt{34}),(atan(\frac{3}{5})),\pi-atan(\frac{3}{5}),\pi+atan(\frac{3}{5}), 2\pi-atan(\frac{3}{5})))$\\
&&& $str(4)=(4,12,(\sqrt{40},\sqrt{40},\sqrt{40},\sqrt{40}),(atan(\frac{2}{6})),\pi-atan(\frac{2}{6}),\pi+atan(\frac{2}{6}), 2\pi-atan(\frac{2}{6})))$\\
&&& $str(5)=(4,9,(\sqrt{18},\sqrt{18},\sqrt{18},\sqrt{18}),(atan(\frac{3}{3})),\pi-atan(\frac{3}{3}),\pi+atan(\frac{3}{3}), 2\pi-atan(\frac{3}{3})))$\\
\hline
2& 6 & 40& $str(1)=(4,12,(\sqrt{25},\sqrt{25},\sqrt{25},\sqrt{25}),(atan(\frac{3}{4})),\pi-atan(\frac{3}{4}),\pi+atan(\frac{3}{4}), 2\pi-atan(\frac{3}{4})))$\\
&&& $str(2)=(4,16,(\sqrt{32},\sqrt{32},\sqrt{32},\sqrt{32}),(atan(\frac{4}{4})),\pi-atan(\frac{4}{4}),\pi+atan(\frac{4}{4}), 2\pi-atan(\frac{4}{4})))$\\
&&& $str(3)=(4,15,(\sqrt{34},\sqrt{34},\sqrt{34},\sqrt{34}),(atan(\frac{3}{5})),\pi-atan(\frac{3}{5}),\pi+atan(\frac{3}{5}), 2\pi-atan(\frac{3}{5})))$\\
&&& $str(4)=(4,20,(\sqrt{41},\sqrt{41},\sqrt{41},\sqrt{41}),(atan(\frac{4}{5})),\pi-atan(\frac{4}{5}),\pi+atan(\frac{4}{5}), 2\pi-atan(\frac{4}{5})))$\\
&&& $str(5)=(4,25,(\sqrt{50},\sqrt{50},\sqrt{50},\sqrt{50}),(atan(\frac{5}{5})),\pi-atan(\frac{5}{5}),\pi+atan(\frac{5}{5}), 2\pi-atan(\frac{5}{5})))$\\
&&& $str(6)=(4,18,(\sqrt{45},\sqrt{45},\sqrt{45},\sqrt{45}),(atan(\frac{3}{6})),\pi-atan(\frac{3}{6}),\pi+atan(\frac{3}{6}), 2\pi-atan(\frac{3}{6})))$\\
\hline
3 & 9 & 40& $str(1)=(4,12,(\sqrt{25},\sqrt{25},\sqrt{25},\sqrt{25}),(atan(\frac{3}{4})),\pi-atan(\frac{3}{4}),\pi+atan(\frac{3}{4}), 2\pi-atan(\frac{3}{4})))$\\
&&& $str(2)=(4,16,(\sqrt{32},\sqrt{32},\sqrt{32},\sqrt{32}),(atan(\frac{4}{4})),\pi-atan(\frac{4}{4}),\pi+atan(\frac{4}{4}), 2\pi-atan(\frac{4}{4})))$\\
&&& $str(3)=(4,15,(\sqrt{34},\sqrt{34},\sqrt{34},\sqrt{34}),(atan(\frac{3}{5})),\pi-atan(\frac{3}{5}),\pi+atan(\frac{3}{5}), 2\pi-atan(\frac{3}{5})))$\\
&&& $str(4)=(4,20,(\sqrt{41},\sqrt{41},\sqrt{41},\sqrt{41}),(atan(\frac{4}{5})),\pi-atan(\frac{4}{5}),\pi+atan(\frac{4}{5}), 2\pi-atan(\frac{4}{5})))$\\
&&& $str(5)=(4,25,(\sqrt{50},\sqrt{50},\sqrt{50},\sqrt{50}),(atan(\frac{5}{5})),\pi-atan(\frac{5}{5}),\pi+atan(\frac{5}{5}), 2\pi-atan(\frac{5}{5})))$\\
&&& $str(6)=(4,12,(\sqrt{40},\sqrt{40},\sqrt{40},\sqrt{40}),(atan(\frac{2}{6})),\pi-atan(\frac{2}{6}),\pi+atan(\frac{2}{6}), 2\pi-atan(\frac{2}{6})))$\\
&&& $str(7)=(4,18,(\sqrt{45},\sqrt{45},\sqrt{45},\sqrt{45}),(atan(\frac{3}{6})),\pi-atan(\frac{3}{6}),\pi+atan(\frac{3}{6}), 2\pi-atan(\frac{3}{6})))$\\
&&& $str(8)=(4,24,(\sqrt{52},\sqrt{52},\sqrt{52},\sqrt{52}),(atan(\frac{4}{6})),\pi-atan(\frac{4}{6}),\pi+atan(\frac{4}{6}), 2\pi-atan(\frac{4}{6})))$\\
&&& $str(9)=(4,30,(\sqrt{61},\sqrt{61},\sqrt{61},\sqrt{61}),(atan(\frac{5}{6})),\pi-atan(\frac{5}{6}),\pi+atan(\frac{5}{6}), 2\pi-atan(\frac{5}{6})))$\\
\hline
4 & 20 & 100& $str(1)=(4,10,(\sqrt{22.25},\sqrt{22.25},\sqrt{22.25},\sqrt{22.25}),(atan(\frac{2.5}{4})),\pi-atan(\frac{2.5}{4}),\pi+atan(\frac{2.5}{4}), $\\
&&& $2\pi-atan(\frac{2.5}{4})))$\\
&&& $str(2)=(4,8,(\sqrt{20},\sqrt{20},\sqrt{20},\sqrt{20}),(atan(\frac{4}{2})),\pi-atan(\frac{4}{2}),\pi+atan(\frac{4}{2}), 2\pi-atan(\frac{4}{2})))$\\
&&& $str(3)=(4,15,(\sqrt{34},\sqrt{34},\sqrt{34},\sqrt{34}),(atan(\frac{3}{5})),\pi-atan(\frac{3}{5}),\pi+atan(\frac{3}{5}), 2\pi-atan(\frac{3}{5})))$\\
&&& $str(4)=(4,14,(\sqrt{28.25},\sqrt{28.25},\sqrt{28.25},\sqrt{28.25}),(atan(\frac{4}{3.5})),
      \pi-atan(\frac{4}{3.5}),\pi+atan(\frac{4}{3.5}), $\\
&&& $2\pi-atan(\frac{4}{3.5})))$\\
&&& $str(5)=(4,7.50,(\sqrt{27.25},\sqrt{27.25},\sqrt{27.25},\sqrt{27.25}),(atan(\frac{1.5}{5})),\pi-atan(\frac{1.5}{5}),\pi+atan(\frac{1.5}{5}),$\\
&&& $ 2\pi-atan(\frac{1.5}{5})))$\\
&&& $str(6)=(4,18,(\sqrt{45},\sqrt{45},\sqrt{45},\sqrt{45}),(atan(\frac{3}{6})),\pi-atan(\frac{3}{6}),\pi+atan(\frac{3}{6}), 2\pi-atan(\frac{3}{6})))$\\
&&& $str(7)=(4,12,(\sqrt{40},\sqrt{40},\sqrt{40},\sqrt{40}),(atan(\frac{2}{6})),\pi-atan(\frac{2}{6}),\pi+atan(\frac{2}{6}), 2\pi-atan(\frac{2}{6})))$\\
&&& $str(8)=(4,18,(\sqrt{45},\sqrt{45},\sqrt{45},\sqrt{45}),(atan(\frac{3}{6})),\pi-atan(\frac{3}{6}),\pi+atan(\frac{3}{6}), 2\pi-atan(\frac{3}{6})))$\\
&&& $str(9)=(4,20,(\sqrt{41},\sqrt{41},\sqrt{41},\sqrt{41}),(atan(\frac{5}{4})),\pi-atan(\frac{5}{4}),\pi+atan(\frac{5}{4}), 2\pi-atan(\frac{5}{4})))$\\
&&& $str(10)=(4,5.25,(\sqrt{14.50},\sqrt{14.50},\sqrt{14.50},\sqrt{14.50}),(atan(\frac{1.5}{3.5})),\pi-atan(\frac{1.5}{3.5}),\pi+atan(\frac{1.5}{3.5}),$\\
&&& $ 2\pi-atan(\frac{1.5}{3.5})))$\\
&&& $str(11)=(4,12,(\sqrt{25},\sqrt{25},\sqrt{25},\sqrt{25}),(atan(\frac{3}{4})),\pi-atan(\frac{3}{4}),\pi+atan(\frac{3}{4}), 2\pi-atan(\frac{3}{4})))$\\
&&& $str(12)=(4,6,(\sqrt{18.25},\sqrt{18.25},\sqrt{18.25},\sqrt{18.25}),(atan(\frac{1.5}{4})),
\pi-atan(\frac{1.5}{4}),\pi+atan(\frac{1.5}{4}),$\\
&&& $ 2\pi-atan(\frac{1.5}{4})))$\\
&&& $str(13)=(4,15,(\sqrt{34},\sqrt{34},\sqrt{34},\sqrt{34}),(atan(\frac{3}{5})),\pi-atan(\frac{3}{5}),\pi+atan(\frac{3}{5}), 2\pi-atan(\frac{3}{5})))$\\
&&& $str(14)=(4,20,(\sqrt{41},\sqrt{41},\sqrt{41},\sqrt{41}),(atan(\frac{4}{5})),\pi-atan(\frac{4}{5}),\pi+atan(\frac{4}{5}), 2\pi-atan(\frac{4}{5})))$\\
&&& $str(15)=(4,17.50,(\sqrt{37.25},\sqrt{37.25},\sqrt{37.25},\sqrt{37.25}),(atan(\frac{3.5}{5})),
\pi-atan(\frac{3.5}{5}),\pi+atan(\frac{3.5}{5}),$\\
&&& $ 2\pi-atan(\frac{3.5}{5})))$\\
&&& $str(16)=(4,15,(\sqrt{42.25},\sqrt{42.25},\sqrt{42.25},\sqrt{42.25}),(atan(\frac{2.5}{6})),\pi-atan(\frac{2.5}{6}),\pi+atan(\frac{2.5}{6}),$\\
&&& $ 2\pi-atan(\frac{2.5}{6})))$\\
&&& $str(17)=(4,12,(\sqrt{40},\sqrt{40},\sqrt{40},\sqrt{40}),(atan(\frac{2}{6})),\pi-atan(\frac{2}{6}),\pi+atan(\frac{2}{6}), 2\pi-atan(\frac{2}{6})))$\\
&&& $str(18)=(4,20,(\sqrt{41},\sqrt{41},\sqrt{41},\sqrt{41}),(atan(\frac{4}{5})),\pi-atan(\frac{4}{5}),\pi+atan(\frac{4}{5}), 2\pi-atan(\frac{4}{5})))$\\
&&& $str(19)=(4,30,(\sqrt{61},\sqrt{61},\sqrt{61},\sqrt{61}),(atan(\frac{5}{6})),\pi-atan(\frac{5}{6}),\pi+atan(\frac{5}{6}), 2\pi-atan(\frac{5}{6})))$\\
&&& $str(20)=(4,9,(\sqrt{18},\sqrt{18},\sqrt{18},\sqrt{18}),(atan(\frac{3}{3})),\pi-atan(\frac{3}{3}),\pi+atan(\frac{3}{3}), 2\pi-atan(\frac{3}{3})))$\\
\hline
\end{tabular}
\label{table_literature_instances}
}
\end{centering}
\end{table}

We run each algorithm 50 times on each instance, recording the best radius found, $r_{best}$, average radius, $\overline{r}$, standard deviation of radii, $r_{\sigma}$ and average run time in seconds, $\overline{t}$. Each algorithm is coded in C, compiled with g++ 4.1.0 , and run under SUSE Linux 10.1 (kernel version 2.6.16.54-0.2.5-smp) on a computer with dual Intel Harpertown E5462 2.80GHz processors, 4GB of RAM and an 80GB hard drive.

The three algorithms (SA, CA-PSLS and PSO) each run over a number of iterations, which is dictated by the value of the constant CYCLE. In this set of experiments, we set CYCLE=3000 for each algorithm. The SA parameter values for $cmax$, initial temperature, $\lambda_ 1$, and $\lambda_ 2$ are set as before, and the $imax$ values set as to 100000, 120000, 108000 and 100000 for instances 1-4 respectively. The results obtained are depicted in Table ~\ref{table_results_new}. 

\begin{table} 
\begin{centering}
\caption{Results for SA, CA-PSLS and PSO on four rectangular instances from the literature (CYCLE=3000)}
\begin{tabular}{c c c c c c c}
\hline
Instance & Size		& Algorithm		& $r_{best}$ 	& $\overline{r}$& $r_{\sigma}$ & $\overline{t}$ (s)\\ \hline

1 & 5        	& SA			& 12.776       	& 13.693       	& 0.62     & 0.82  \\
&		& CA-PSLS	    	& 10.942       	& 11.704       	& 0.49	   & 4.31  \\	
&		& PSO			& 11.046       	& 11.716       	& 0.49	   & 4.89  \\
\hline		
2 & 6		& SA			& 16.004        & 17.377       	& 0.69	   & 1.66  \\
&		& CA-PSLS	    	& 14.686        & 15.590       	& 0.67 	   & 7.76  \\
&		& PSO 			& 14.320        & 15.349        & 0.56     & 8.28  \\
\hline		
3 & 9		& SA			& 20.849        & 22.328        & 0.87	   & 6.84  \\ 
&		& CA-PSLS	    	& 18.157        & 19.797        & 1.03	   & 21.07 \\
&		& PSO			& 18.579        & 19.205        & 0.49	   & 22.84 \\
\hline	
4 & 20		& SA			& 29.969        & 31.680       	& 0.98	   & 92.01 \\
&		& CA-PSLS	    	& 27.927        & 33.129        & 5.48 	   & 125.52\\
&		& PSO			& 32.596        & 34.426        & 2.23	   & 138.00\\
\\ \hline
\end{tabular}
\label{table_results_new}
\end{centering}
\end{table}

On the first three (small) instances, both particle-based algorithms slightly out-perform the SA method in terms of solution {\it quality}; on average, by 10\%. However, this comes at a significant cost disadvantage in terms of run time; over the first three instances, the particle-based methods require four times the execution time of the SA algorithm to terminate. When the problem size is increased to 20, the benefits of the SA algorithm begin to become apparent, as it out-performs the other two algorithms in terms of both solution quality {\it and} run time.
In order to establish the significance of this, we now test all three methods on much larger instances.

\subsection{Large Rectangular Instances}

\begin{figure}[h]
\begin{centering}
\includegraphics[scale=1.1]{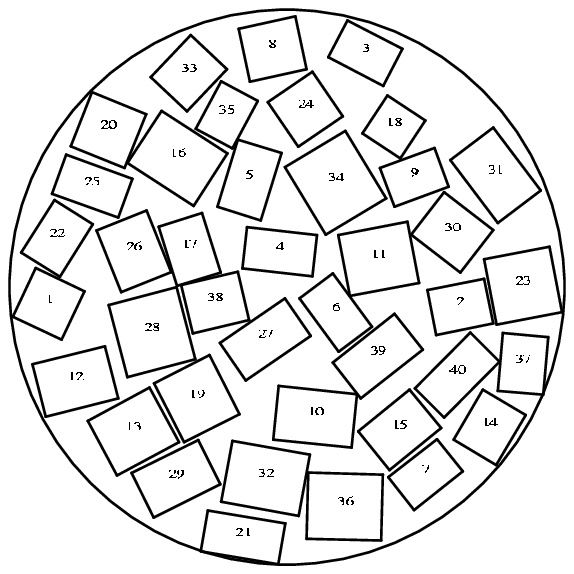}
\caption{Best 40 rectangle solution generated by SA ($r=134.07$)}
\label{fig:40}
\end{centering}
\end{figure}

We designed instances with 40, 60, 80 and 100 rectangles. Space precludes a detailed description of these, but the full problem set is available from the corresponding author. As before, each method was run 50 times on each instance. Because of the computational cost incurred, we reduced CYCLE to 1000 for each algorithm. The results are depicted in Table ~\ref{table_results_big}, with an example solution for the 40 rectangle instance depicted in Figure ~\ref{fig:40}.

\begin{table}[h] 
\begin{centering}
\caption{Results for SA, CA-PSLS and PSO on large instances (CYCLE=1000)}
\begin{tabular}{c c c c c c c}
\hline
Instance & Size		& Algorithm		 & $r_{best}$ 	& $\overline{r}$& $r_{\sigma}$ & $\overline{t}$ (s)\\ \hline

1 & 40        	& SA			& 164.061      	& 174.586      	& 4.81 	   & 263.24   \\
&		& CA-PSLS	    	& 179.508      	& 253.627      	& 60.42    & 219.84   \\	
&		& PSO			& 242.471     	& 276.939      	& 24.44    & 197.03   \\
\hline		
2 & 60		& SA			& 170.284       & 187.312      	& 6.32 	   & 905.75   \\
&		& CA-PSLS	    	& 184.984       & 288.642      	& 124.26   & 579.31   \\
&		& PSO 			& 272.282       & 317.739       & 26.17	   & 451.72   \\
\hline		
3 & 80		& SA			& 265.654       & 281.087       & 8.54 	   & 2178.31  \\ 
&		& CA-PSLS	    	& 298.524       & 544.421       & 162.81   & 1016.70  \\
&		& PSO			& 432.347       & 490.862       & 35.75	   & 813.59   \\
\hline	
4 & 100		& SA			& 406.991       & 423.087    	& 7.87 	   & 4260.54  \\
&		& CA-PSLS	    	& 658.352       & 880.537       & 108.65   & 1611.52  \\
&		& PSO			& 598.265       & 688.785       & 48.06	   & 1277.43  \\
\\ \hline
\end{tabular}
\label{table_results_big}
\end{centering}
\end{table}

The SA method significantly out-performs the other two methods in terms of solution quality, but with an associated cost in terms of run time. However, as shown by the figures for standard deviation, SA offers a consistently high-quality solution method (at a price), whereas the other two algorithms offer solutions of more variable quality, but more quickly. 


\section{Conclusions}

In this paper we describe a novel algorithm based on simulated annealing for the problem of packing weighted polygons inside a circular container. As well as being of significant theoretical interest, this problem has real significance in domains such as satellite design in the aerospace industry. Our algorithm consistently generates high-quality solutions that offer a significant improvement over those generated by other methods. However, this superiority comes with an associated computational overhead, so the choice of method should largely be driven by the anticipated application. Future work will involve improving the method's performance on problems containing nonconvex polygons, as well as its extension into three dimensions.

\section*{Acknowledgements}

This work was partially supported by the Dalton Research Institute, Manchester Metropolitan University.

\bibliography{packing}

\end{document}